# A mechanically strong and ductile soft magnet with extremely low coercivity

Liuliu Han[1], Fernando Maccari[2], Isnaldi R. Souza Filho[1], Nicolas J. Peter[1], Ye Wei[1], Baptiste Gault[1], Oliver Gutfleisch[2], Zhiming Li[3], Dierk Raabe[1]

[1]*Max-Planck-Institut für Eisenforschung, Max-Planck-Straße 1, 40237 Düsseldorf, Germany*

[2]*Department of Material Science, Technical University of Darmstadt, 64287 Darmstadt, Germany*

[3]*School of Materials Science and Engineering, Central South University, 410083 Changsha, China*

*Correspondence to lizhiming@csu.edu.cn (Z.L.), d.raabe@mpie.de (D.R.)

Soft magnetic materials (SMMs) serve in electrical applications and sustainable energy supply, allowing magnetic flux variation in response to changes in applied magnetic field, at low energy loss[1]. The electrification of transport, households and manufacturing leads to an increase in energy consumption due to hysteresis losses[2]. Therefore, minimizing coercivity, which scales these losses, is crucial[3]. Yet, meeting this target alone is not enough: SMMs in electrical engines must withstand severe mechanical loads, i.e., the alloys need high strength and ductility[4]. This is a fundamental design challenge, as most methods that enhance strength introduce stress fields that can pin magnetic domains, thus increasing coercivity and hysteretic losses[5]. Here, we introduce an approach to overcome this dilemma. We have designed a Fe-Co-Ni-Ta-Al multicomponent alloy with ferromagnetic matrix and paramagnetic coherent nanoparticles (~91 nm size, ~55% volume fraction). They impede dislocation motion, enhancing strength and ductility. Their small size, low coherency and small magnetostatic energy create an interaction volume below the magnetic domain wall width, leading to minimal domain wall pinning, thus maintaining the soft magnetic properties. The alloy has a tensile strength of 1336 MPa at 54% tensile elongation, extremely low coercivity of 78 A/m (<1 Oe), moderate saturation magnetization of 100 Am$^2$/kg, and high electrical resistivity of 103 μΩ·cm.

Lowest possible coercivity and highest possible electrical resistivity are primary goals for soft magnetic materials (SMMs), to reduce hysteresis- and eddy current-related energy losses, noise and the associated material damage[1-3]. Also, new SMMs with higher strength and ductility are needed, to operate under mechanically demanding loading conditions for safety-critical parts in transportation and energy[4]. High strength and ductility serve also as measures for many other mechanical properties, such as high hardness[5] and fracture toughness[6]. This multi-property profile creates a fundamental dilemma. Mechanical strength of metallic materials is produced by lattice defects and their elastic interactions with linear lattice faults that carry inelastic deformation, referred to as dislocations. However, the defects also interact with the magnetic domain walls and pin them. The loss in domain wall motion increases coercivity so that the materials lose their soft magnetic features. Therefore, current SMMs follow the design rule of avoiding lattice defects to minimize coercivity[7]. Vice versa, increasing an alloy's mechanical strength requires enhancing its internal stress level through defects such as dislocations, grain boundaries and precipitates[8]. This means that the task of making soft magnets mechanically strong is a trade-off between two mutually exclusive design strategies, namely, mechanical strength vs. unaffected domain wall motion.

The theory of the grain size dependence of coercivity[9] shows its proportionality to the 6$^{th}$ power of the grain size for the case of nanocrystalline materials, a relation that can also be applied to particles[10]. Current design of SMMs has thus focused on using small particles (<15 nm)[10,11] and grain sizes (<100 nm)[12-14]. According to magnetic strain theory, the coercivity depends on the energy required to displace domain walls to overcome lattice barriers[15-17]. Here we introduce particles into a multicomponent massive solid solution matrix and increase their size from the commonly used range of 5–15 nm to 90–100 nm. With that, the internal stress level and the overall elastic coherency misfit energy are reduced through the smaller specific surface area (total surface area per unit of volume) of the particles, caused by the coarsening. We then propose that the particle design must follow four main rules. First, minimal pinning of domain walls requires a well-tuned and -controlled particle size distribution with optimum balance between the decrease in specific surface area and the increase in magnetostatic energy during particle coarsening. Second, the particle size must remain smaller than the domain wall width to prevent strong pinning, i.e. strong resistance against spin rotation[8]. Third, the particles' chemical composition and crystal structure determine their saturation magnetization; therefore, antiferromagnetic elements are usually excluded. Fourth, strengthening of the alloys is determined by the interaction between dislocations and particles and by the friction forces exerted on dislocations in the massive solid solution matrix. Thus, intrinsically strong intermetallic particles with minimal lattice misfit are targetted. These require high forces for dislocation cutting (providing strength), but repeated cutting by ensuing dislocations emitted by the same source shear them with gradual ease along the remaining and gradually reducing particle cross sections (providing ductility).

These different mechanism considerations had next to be translated into a corresponding compositional alloy design concept. This is mainly guided by the requirement for (a) a ferromagnetic matrix with (b) high solid solution contribution, and components that trigger the formation of (c) strong and stable intermetallic phases with (d) small lattice misfit relative to the matrix. These considerations have led us to the non-equiatomic iron–nickel–cobalt–tantalum–aluminium ($Co_{27.7}Fe_{32.6}Ni_{27.7}Ta_{5.0}Al_{7.0}$, at.%) multicomponent alloy (MCA). We synthesized the material in a vacuum induction melting furnace followed by conventional hot rolling and homogenization (details of the processing procedures and chemical compositions are provided in the Methods section). Through further isothermal heat treatments (1–100 h at 1173 K), we prepared samples with different average particle sizes, ranging from 24 ± 15 nm to 255 ± 49 nm (edge length is used to characterize the topological particle size). The particles have $L1_2$ structure and complex composition, as presented in detail below.

Fig. 1 shows the structural characterization of the MCA with medium particle size (M-MCA, where "M" stands for "medium particle size") after annealing at 1173 K for 5 h. The M-MCA shows an average grain size of 85.3 ± 25.6 μm (excluding annealing twin boundaries) according to electron backscatter diffraction (EBSD) analysis shown in Fig. 1a. The electron channelling contrast imaging (ECCI) analysis shows that the $L1_2$ particles have a high number density (7.2 ± 0.2)×$10^{20}$ m$^{-3}$ and a large volume fraction (55% ± 1%) in homogeneous distribution within the face-centred cubic (fcc) matrix (Fig. 1b). The lattice misfit (0.48%) between the fcc matrix and the $L1_2$ particles has been calculated using their lattice parameters[18] acquired from the X-ray diffraction (XRD) patterns (Fig. 1c) by Rietveld simulation[19]. Such a small lattice mismatch reduces the capillary driving forces for further coarsening and the uniform dispersion prevents plastic localization at high strength[20,21]. The central beam dark-field (DF) transmission electron microscopy (TEM) analysis reveals that the average size of the L1$_2$ particles is 90.8 ± 35.8 nm, Fig. 1d. The corresponding selected-area electron diffraction (SAED, see inset of Fig. 1d) and high-resolution (HR)-TEM (Extended Data Fig. 1d) confirm the high coherency between the particles and the matrix.

The elemental partitioning between the L1$_2$ precipitates and fcc solid solution matrix is characterized by atom probe tomography (APT). Fig. 1e shows the three-dimensional (3D) distribution of the volume investigated by APT, highlighted by a set of iso-surfaces delineating regions containing above 25 at.% Fe. Fig. 1f shows the one-dimensional (1D) compositional profiles acquired along the cylinder in Fig. 1e. The profiles show that Fe partitions into the fcc matrix (36 at.%), whereas the L1$_2$ particles are enriched in Ni (40 at.%), Ta (13 at.%), and Al (9 at.%). The compositions of the fcc and L1$_2$ phases were determined by averaging over 3 APT data sets (including 10 L1$_2$ particles) as Fe$_{36}$Co$_{28}$Ni$_{26}$Al$_7$Ta$_3$ and Ni$_{40}$Co$_{26}$Ta$_{13}$Fe$_{12}$Al$_9$ (at.%), respectively.

Besides these intragranular nanoparticles, we also observed two types of grain-boundary variants: (a) Coarse grain boundary particles (160.2 ± 55.3 nm) with the same crystal structure and composition as the ones inside the grains (Extended Data Fig. 1a-c), and (b) incoherent particles with a minor fraction (<0.3%) at the triple-points of the grains with different structure (cubic Fd-3m, Fig. 1c) and composition (Ta$_{40}$Co$_{26}$Fe$_{20}$Ni$_{11}$Al$_3$, at.%, Extended Data Fig. 1e). These two types of particles are promoted by the high diffusion rate along the grain boundaries.

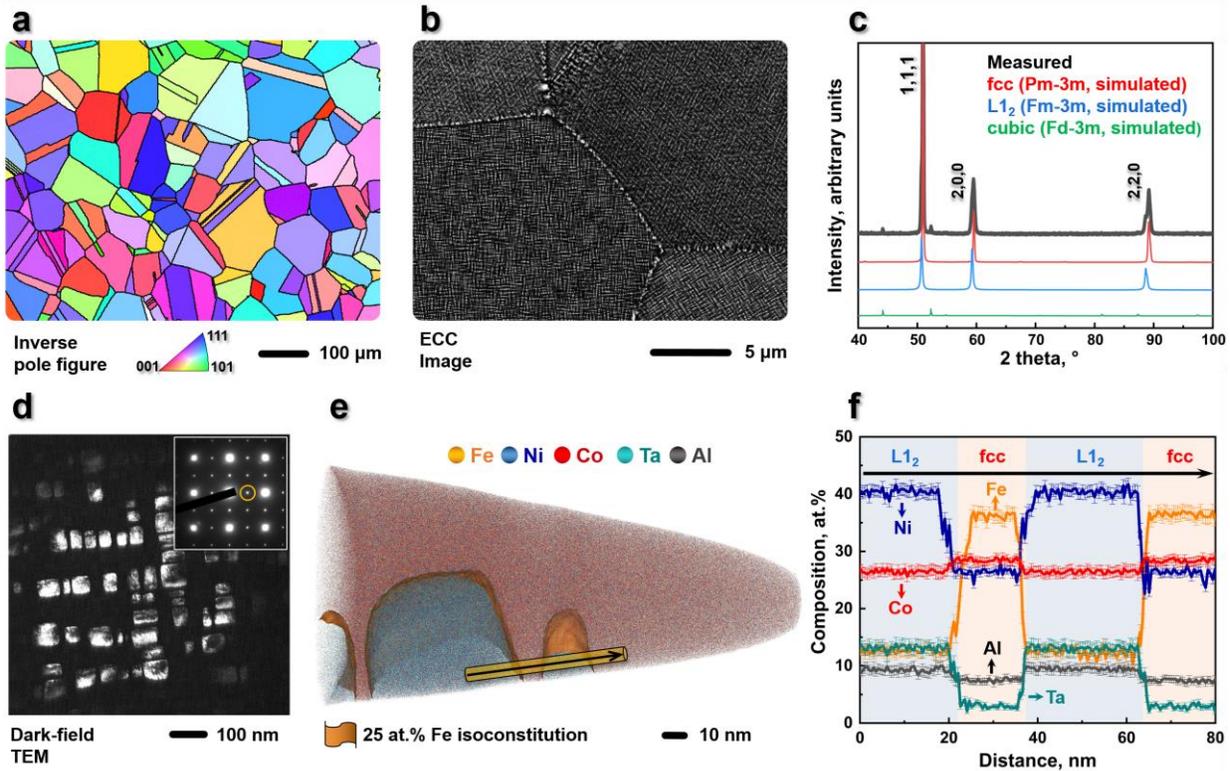

**Figure 1 | Microstructure and chemical composition of the M-MCA. a,** EBSD-inverse pole figure (IPF) map showing the equiaxed grains of the fcc matrix. The black lines highlight the high-angle grain/twin boundaries. **b,** ECC image featuring the high-density uniformly distributed L1$_2$ particles in the grain interiors and heterogeneous particles at the grain boundaries. **c,** Measured and simulated XRD patterns showing the phase structures. **d,** Centered dark-field TEM image of the L1$_2$ particles obtained using the (011) superlattice spot (see inset). **e,** 3D reconstruction map of a typical APT tip showing the cuboidal L1$_2$ particles embedded in the fcc matrix. The L1$_2$/fcc interfaces are highlighted using iso-composition surfaces containing 25 at.% Fe. **f,** 1D compositional profiles computed along the cylinder region (marked by black arrow) in **e**, showing the compositional changes across several interfaces. Error bars refer to the standard deviations of the counting statistics in each bin of the profiles.

The current strategy of tuning the particle size also allows overcoming the strength-ductility trade-off, significant for advanced alloys with Giga-Pascal-level strength. Fig. 2a displays the tensile stress-strain curves of the M-MCA at room–temperature (red curve). The yield strength ($\sigma_y$) is 904 ± 11 MPa with an ultimate tensile strength ($\sigma_{UTS}$) of 1336 ± 21 MPa, and an elongation at fracture ($\varepsilon_f$) up to 53.6 ± 1.5%, averaged from four tests. Accordingly, the M-MCA has a high $\sigma_{UTS}\times\varepsilon_f$ value of 71.6 GPa%. To reveal the improvement in strength and ductility achieved by the well-controlled particle size distribution, the mechanical response of a material variant with identical chemical composition, i.e., $Fe_{32}Co_{28}Ni_{28}Ta_5Al_7$ (at.%), but smaller particle size (S-MCA, annealed for 1 h, producing an average particle size of 24 nm), larger particle size (L-MCA, annealed for 100 h, average particle size of 255 nm) and the particle-free $Fe_{35}Co_{30}Ni_{30}Ta_5$ (at.%) alloy[11], are also presented in Fig. 2a.

Compared with the single-phase $Fe_{35}Co_{30}Ni_{30}Ta_5$ (at.%) alloy with a relatively low value of $\sigma_y$ of 501 MPa[11], the notable increase in yield strength of the M-MCA can be attributed to the precipitation strengthening of the $L1_2$ particles with a high volume fraction (55%). More importantly, such improvement in the strength of the M-MCA is achieved at no expense of ductility, which is fundamentally different from the case of the S-MCA where a significant loss in ductility is observed with increasing strength. The good ductility is correlated with the high work-hardening capability, as shown in Fig. 2b. Further increase of particle size at longer annealing time (100 h) in the L-MCA leads to the decrease of ductility (53%) and $\sigma_{UTS}$ (14%) as compared to those of the M-MCA. This is related to the mechanical weakness, strain localization and embrittlement in the particle-free zone adjacent to the grain boundary caused by solute depletion (Extended Data Fig. 2a). This interfacial weakening has been confirmed by the associated fracture morphologies, with typical ductile fracture with fine dimples in the M-MCA material (see Fig. 2b inset) and intergranular fracture in the L-MCA material (Extended Data Fig. 2b).

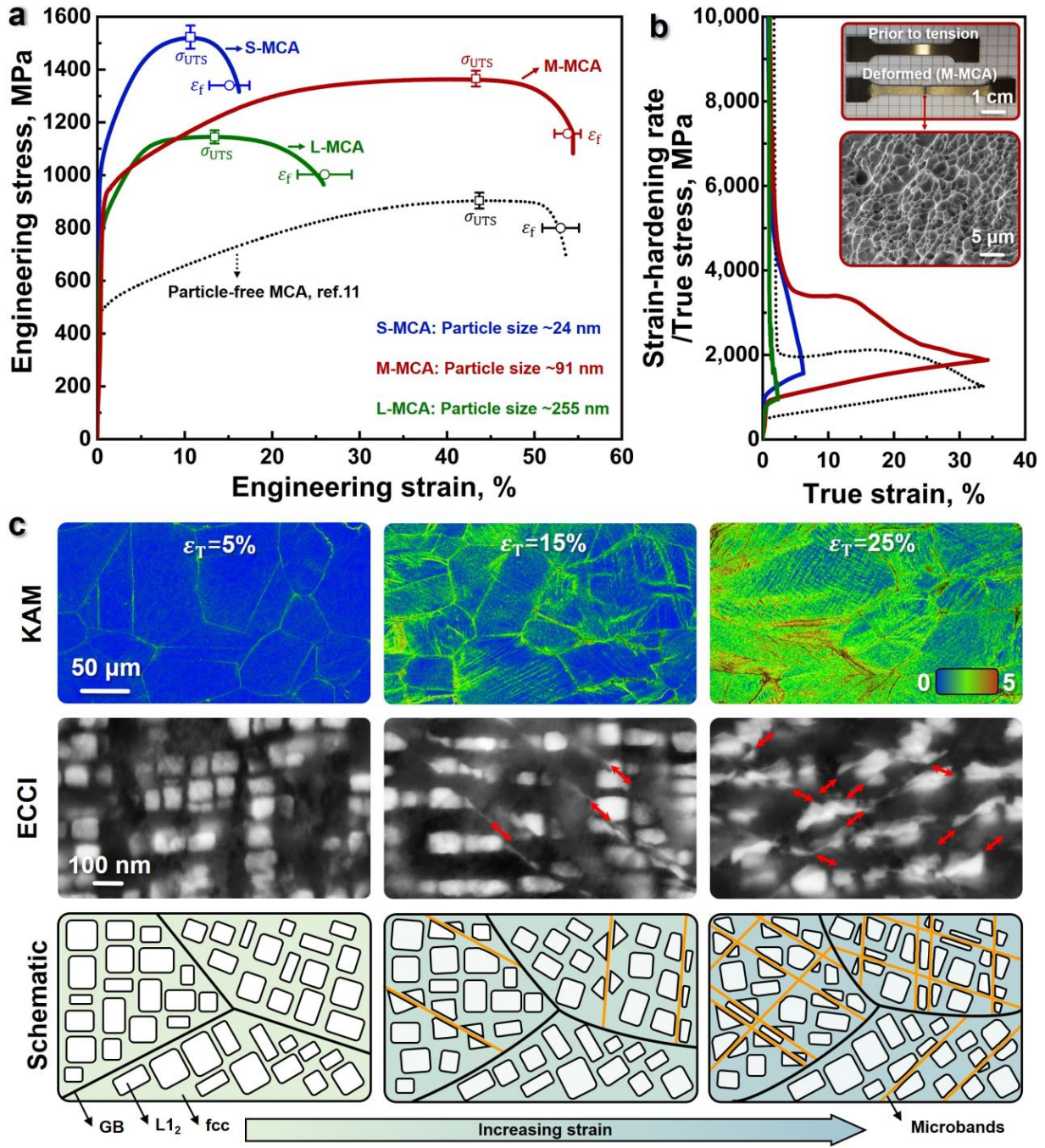

**Figure 2 | Mechanical behaviour and nanoscale processes during plastic straining of the M-MCA. a,** Typical engineering stress-strain curves measured at room–temperature together with the average values for ultimate tensile strength ($\sigma_{UTS}$) and elongation at fracture ($\varepsilon_f$). **b,** Strain hardening rate/true stress–true strain curves. The insets show the macroscopic image (upper inset) of the tensile sample and the corresponding fracture morphology (bottom inset), where a typical

ductile fracture with fine dimples is observed. **c,** Substructure evolution as a function of global strain observed after interrupted tensile tests: EBSD–KAM maps revealing the distributions of deformation-induced misorientations (upper), where $\varepsilon_T$ stands for the global true strain; ECCI analysis (middle) showing the evolution of microbands. The shearing of L1$_2$ particles is highlighted by red arrows; schematics (bottom) illustrating the microband refinement in the M-MCA during plastic straining.

To unravel the mechanisms responsible for the significant improvement in the strength-ductility combination of the M-MCA, we revealed its deformation substructures using EBSD kernel average misorientation (KAM) analysis and ECCI at different hardening stages (Fig. 2c). In principle, high strength requires to impede dislocation movement while good ductility needs mobility of dislocations and new production of dislocations[22]. At the early deformation stage, the M-MCA deforms by planar dislocation glide on {111} planes (Extended Data Fig. 3a-c), as commonly observed in fcc alloys[23]. The dislocations extend through the grains. Corresponding pile-up configurations at the grain boundaries are revealed by the higher KAM values (e.g., at $\varepsilon_T = 5\%$; see KAM maps in Fig. 2c). The relatively large grain size (85.3 μm) of the current M-MCA enables higher mobility of dislocations compared to most of the previously reported strong MCAs that had smaller grain size[24-26] (around 10 μm). Further straining refines the crystallographically aligned microbands and facilitates the shearing of L1$_2$ particles (e.g., at $\varepsilon_T = 15\%$; see ECC images in Fig. 2c). The quantification of the evolution of the average microband spacing reveals a microband refinement process during straining (Extended Data Fig. 3d). The gradually reduced microband spacing causes higher passing stress and thus enhanced strain hardening. This has been proposed to explain the good strength-ductility combinations in some high manganese steels[27-29] and MCAs[30]. Hence, the observed dynamic microband refinement and particle shearing are the prevalent deformation mechanisms in the current MCAs. No Orowan looping was observed, even when increasing the average particle size up to 255 nm for the L-MCA (Extended Data Fig. 2c), as the average particle spacing remains far below the critical value (3094 nm, see Methods) for the activation of dislocation bowing around particles, a mechanism referred to as Orowan effect. Furthermore, the stress required for shearing particles in the M-MCA with a medium particle size (91 nm) with high volume fraction (55%) is 2.2 times larger than that for the S-MCA with a smaller particle size (24 nm) with low volume fraction (43% ± 1%) (see Extended

Data Table 1). Therefore, the high critical shear stress required for cutting the $L1_2$ particles and the dynamic microband refinement during plastic deformation lead to the strong strain hardening capacity of the M-MCA.

Fig. 3a and b show the magnetic properties of the MCAs. All alloys exhibit typical soft ferromagnetic behaviour. The M-MCA shows an excellent combination of extremely low coercivity ($H_c$) of 78 ± 3 A/m (<1 Oe) and moderate saturation magnetization ($M_s$) of 100.2 ± 0.2 $Am^2$/kg. We identified a higher $M_s$ for the alloy variant with larger average particle size (see inset in Fig. 3a). The reason behind this is the change in intrinsic magnetic behaviour, as indicated by the higher Curie temperature ($T_c$), revealed by the thermomagnetic curves (Extended Data Fig. 4a). Two distinct changes of the slope, indicating two ferromagnetic phases, are observed in the S-MCA. In contrast, only one sharp drop is observed in the M-MCA and L-MCA materials, indicating only one ferromagnetic phase. This is further confirmed by measuring the magnetic behaviour of the MCAs at elevated–temperatures (Extended Data Fig. 4b). Considering that both, the fcc and $L1_2$ phases contain high concentrations of ferromagnetic elements, we investigated their individual magnetic response through casting both phases as separate bulk samples with their respective nominal compositions acquired before from APT analysis (see Methods for details). The results reveal that the $L1_2$ bulk phase is paramagnetic while the fcc matrix is ferromagnetic in the M-MCA (Extended Data Fig. 4c). Due to different partitioning, the magnetic behaviour of the $L1_2$ phase varies from ferromagnetic in the S-MCA material variant to paramagnetic in the M-MCA and L-MCA materials. The mechanism behind this transition is the change in the intrinsic spin alignment, which is related to the change in chemical composition (Extended Data Fig. 5) and ordering during annealing. The overall increase in saturation magnetization of the MCAs as a function of particle coarsening is attributed to the change in fcc matrix composition due to elemental partitioning (Extended Data Fig. 5), i.e., specifically to the resultant higher concentration of (Fe+Co) in the fcc matrix. This effect enhances the total average magnetic moment per formula unit and leads to a higher $M_s$.

To gain further insight into the mechanism behind the magnetic response of the M-MCA, we investigated the domain structure using magneto-optical Kerr effect (MOKE) microscopy (Fig. 3c) under different applied magnetic field strengths. Starting from the AC demagnetized state to an applied field of 40 kA/m, the nucleation of magnetic domains is uniformly distributed within the grain. Further increasing the applied field (155 kA/m) leads to domain wall movement and growth of energetically favourable domains. The domains grow unaffected inside the grains but get impeded at grain and twin boundaries (Extended Data Fig. 6). Fig. 3d summarized the statistically averaged particle size distribution with respect to the coercivity of all the MCA samples at different annealed states. The data is acquired by developing an automated processing protocol, as shown in Extended Data Fig. 7. The coercivity first decreases from 763 A/m (S-MCA, average particle size 24 nm) to 78 A/m (M-MCA, average particle size 91 nm) and then increases to 1745 A/m (L-MCA, average particle size 255 nm). Both the average particle size and the grain size increase monotonously with increasing annealing time (Extended Data Table 1). Since the grain size of the MCAs material is above the critical single-domain size, its coercivity decreases with grain coarsening, following the model for the grain size dependence of the coercivity as $H_c \propto 1/D$ (where $D$ is the grain size)[13]. However, the magnitude of the decrease in coercivity due to grain coarsening according to the model is negligible compared to the experimentally observed values: The difference according to the model calculation between the S-MCA and M-MCA material variants is 2 A/m, but the experimentally observed difference is 775 A/m. Accordingly, the energy required for the irreversible displacement of domain walls within the grain is the determining effect for the extremely low coercivity.

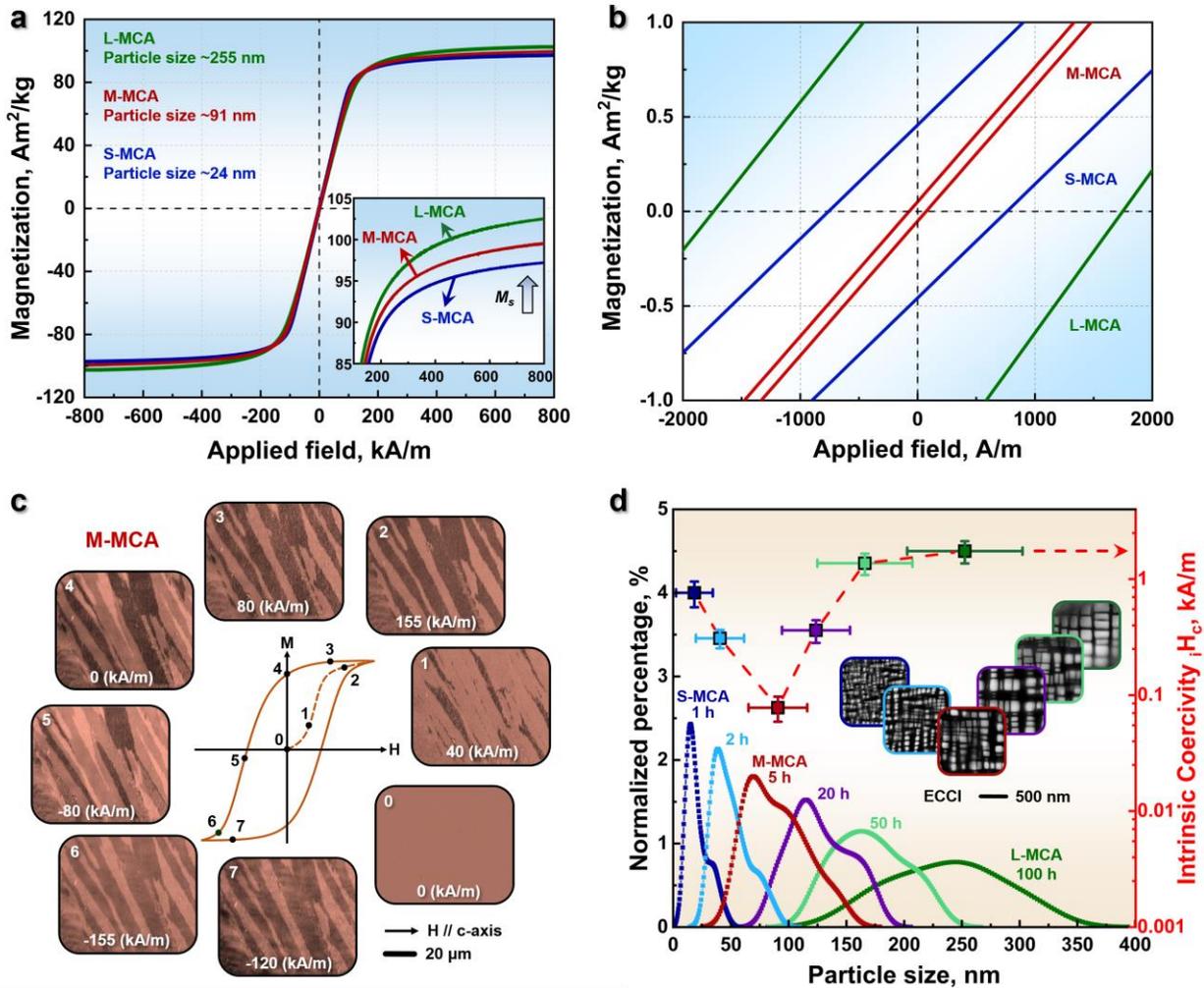

**Figure 3 | Soft magnetic response and associated Bloch wall motion behaviour of the MCAs at room–temperature. a,** Hysteresis loops (*M/H*) acquired up to ± 800 kA/m. The magnetic field-sweeping rate is 1 kA/m. The inset shows the enlarged view of the increase in saturation magnetization with particle coarsening. **b,** *M/H* curves measured at a rate of 0.1 kA/m between ± 50 kA/m, showing the extremely low coercivity. **c,** MOKE microscopy observation, in longitudinal contrast, revealing the magnetization process. The applied magnetic field is horizontal to the viewing plane. **d,** Statistical particle size distribution of all the MCAs under different isothermal heat treatment conditions (1–100 h at 1173 K). The right inset shows the evolution of particle size by ECCI probing.

Therefore, the significant drop in coercivity at the early particle coarsening stage (≤91 nm) is attributed to the gradual reduction of the overall coherency stresses between matrix and particles, due to their average size increase. More specifically, the values of the specific surface area×integrated lattice misfit decrease from $1.09×10^6$ $m^{-1}$ in the S-MCA material to $4.08×10^5$ $m^{-1}$ in the M-MCA material. The dislocation density in the matrix decreases from $1.50×10^{14}$ $m^{-2}$ in the S-MCA material to $9.32×10^{13}$ $m^{-2}$ in the M-MCA material when increasing the annealing time from 1 h to 5 h (see Methods). The reduction in dislocation density lowers the associated elastic distortion fields that can pin the domain walls. Although the elastic distortion and dislocation density decrease with particle coarsening, the coercivity of the L-MCA material increases. Two main mechanisms are proposed to explain this. First, the average particle size and the associated strain field in the L-MCA becomes larger than the domain wall width ($\delta_w$), leading to domain wall pinning. The L-MCA has a $\delta_w$ of 117 nm and a much larger average particle size of 255 nm, whereas the average particle size of the M-MCA is 91 nm, which is below its $\delta_w$ of 112 nm (see Methods). Second, the increased magnetostatic energy ($E_s$) associated with the paramagnetic particles causes a stronger individual pinning effect of each particle on the domain wall motion. More specifically, the $E_s$ of the L-MCA is estimated to be 23 times larger than that of the M-MCA (see Methods). When only considering particle size, it should be noted that for the M-MCA, the coherent particles distributed along the grain boundaries with an average size (160 nm) above the $\delta_w$ (112 nm) are expected to have a stronger pinning effect on the domain walls than those in the grain interiors (91 nm). However, these coarser particles only occupy a small fraction (1.2% ± 0.2% in M-MCA), hence, with a negligible effect on domain wall motion.

These considerations show that the nanoscale size distribution of the particles must be carefully controlled to minimize their pinning effect on domain wall movement, which determines the alloy's coercivity. This is achieved here by an optimal balance between the release of structural defects (e.g., interfacial elastic distortion, dislocation density) just down to a level required to maintain high mechanical strength and the rise of the pinning effect from the magnetostatic energy, while keeping the particle size below the domain wall width during particle coarsening.

To highlight the good combination of mechanical and magnetic properties of the M-MCA with optimal particle size, we compare it with existing SMMs in an Ashby plot showing the $\sigma_{UTS}×\varepsilon_f$ values against the $H_c$, Fig. 4a. This comparison shows that the $\sigma_{UTS} × \varepsilon_f$ value of the new

M-MCA material outperforms all other SMMs. More importantly, the $H_c$ of the new material is lower than that of all Fe-Ni[31,32] alloys and other MCAs[33-44], comparable to that of Fe-Si[45], Fe-Co[46,47] alloys and pure Fe[48]. Amorphous and nanocrystalline soft magnetic alloys[49-52] can exhibit ultralow $H_c$ (<10 A/m) and high mechanical strength, yet, their limited ductility, damage tolerance and workability prohibit their application in cases where load path changes, high stresses, forming or machining are applied. A radar plot comparing the various soft magnetic and mechanical properties of the current M-MCA with several typical commercial SMMs is shown in Extended Data Fig. 8. Although the saturation induction ($B_s$) of the current M-MCA is not comparable to that of typical commercial SMMs (Extended Data Fig. 9), it has higher electrical resistivity ($\rho$) (103 ± 0.8 μΩ·cm, see Extended Data Fig. 10), a feature that makes it suited for applications with alternating current conditions. The high $\rho$ of the M-MCA is expected to be derived from the high resistance to electron movement from the larger lattice distortion[53]. Fig. 4b compares the $\sigma_{UTS} \times \varepsilon_f$ vaues versus the grain size of the M-MCA material with recently reported strong and ductile MCAs[21,24-26,54-57]. The analysis shows that the current alloy reaches high values of $\sigma_{UTS} \times \varepsilon_f$ even without the substantial contribution from grain boundaries, confirming the significant strengthening effect provided by the nanoparticles and the massive solid solution.

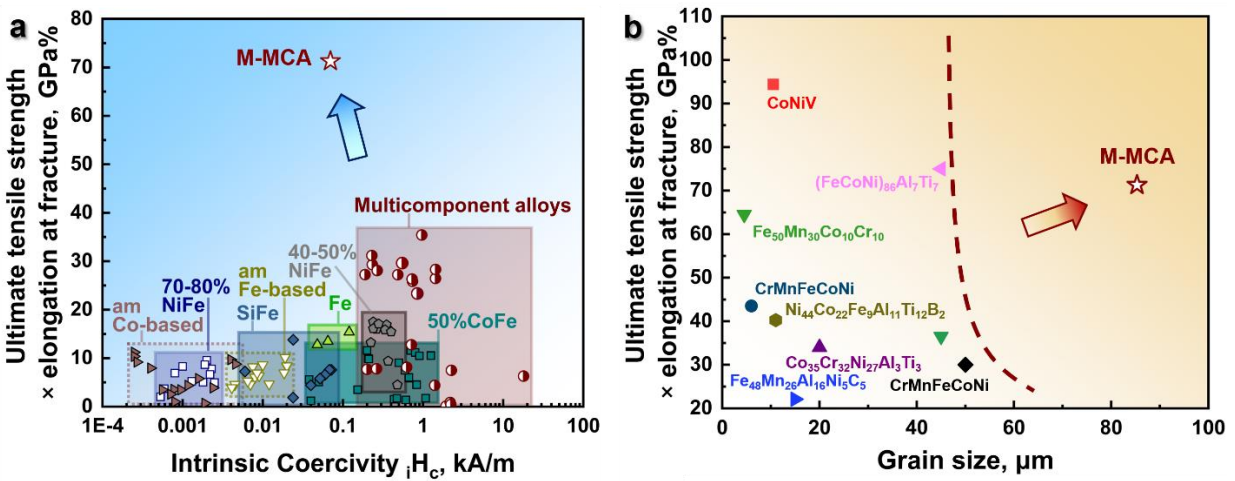

**Figure 4 | Mechanical and magnetic features combined in the new $Fe_{32}Co_{28}Ni_{28}Ta_5Al_7$ (at.%) M-MCA material. a,** Ashby map compiling room–temperature ultimate tensile strength ($\sigma_{UTS}$)×elongation at fracture ($\varepsilon_f$) and intrinsic coercivity compared to other soft magnetic materials, such as Fe-Ni[31,32], Fe-Si[45], Fe-Co[46,47], Fe[48], amorphous alloys[49-52] and established

MCAs[33-44]. **b,** Ashby map showing $\sigma_{UTS} \times \varepsilon_f$ versus average grain size compared to those of other strong and ductile MCAs[21,24-26,54-57]. 'am' in the legend stands for amorphous alloys.

In summary, we have developed a material that unifies so far mutually exclusive properties, namely, high mechanical strength (1336 MPa), high tensile ductility (54%), low coercivity (78 A/m), moderate saturation magnetization (100 Am$^2$/kg) and high resistivity (103 μΩ·cm). We realized this in a new class of bulk SMMs through a nanoparticle dispersion with well-controlled size (91 nm), magnetic properties, coherency strain, strength and interface energy. The design strategy is opposite to that generally applied in conventional SMM design. Instead of using smallest microstructure features (particle size <15 nm) to avoid magnetic wall pinning as in conventional SMMs, we have chosen a relatively coarse particle dispersion with tuned particle/matrix interfacial coherency stresses and paramagnetic properties to minimize magnetic pinning of domain walls on the one hand (soft magnetism) and maximize interaction strength with dislocations on the other hand (strength and ductility).

The infinite composition space of MCAs allows to realize materials with good combinations of soft magnetic and mechanical properties. The new alloy design approach allows for tailoring SMMs for magnetic parts exposed to severe mechanical loads, be it from manufacturing and/or during service, for which conventional SMMs are mechanically too soft or too brittle[53]. Future efforts on developing advanced magnetic MCAs could target variants with further improved soft magnetic properties (e.g., higher magnetic saturation) while preserving their outstanding mechanical properties, at lower alloy costs, and use high throughout experiments combined with computational techniques, e.g., machine learning[58], to accelerate the discovery of new alloy variants.

**Methods**

**Materials preparation.** The bulk MCA ingot with a predetermined nominal composition of Co$_{27.7}$Fe$_{32.6}$Ni$_{27.7}$Ta$_{5.0}$Al$_{7.0}$ (at%) was first cast in a vacuum induction furnace using pure metallic ingredients (purity higher than 99.8 wt.%) under high-purity argon (Ar) atmosphere. The as-cast ingot with a dimension of 40 mm×60 mm×20 mm (*length×width×thickness*) was then hot-rolled at 1473 K to an engineering thickness reduction of 50% (final thickness 10 mm). After hot-rolling, The alloy sheets were then homogenized at 1473 K for 10 min in Ar atmosphere followed by water-quenching. To obtain a wide size distribution of the particles, further isothermal heat treatments were conducted at 1173 K, lasting from 1 h up to 100 h (1 h, 2 h, 5 h, 20 h, 50 h, 100 h) in Ar atmosphere and followed by water quenching. The exact chemical composition of the MCA measured by wet-chemical analysis is $Co_{28.0}Fe_{32.0}Ni_{28.1}Ta_{4.7}Al_{7.2}$ (at.%), which is close to the pre-designed composition. In addition, the bulk ingots (50 g) with the compositions identical to that of the fcc ($Fe_{36}Co_{28}Ni_{26}Al_7Ta_3$, at%) matrix phase and the $L1_2$ particle ($Ni_{40}Co_{26}Ta_{13}Fe_{12}Al_9$, at.%) in the M-MCA derived from the APT analysis were also cast, respectively, by arc-melting under Ar atmosphere. The ingots were remelted six times to achieve chemical homogeneity.

**Analytical methods.** X-ray diffraction (XRD) measurements were carried out in an X-ray system (Diffractometer D8 Advance A25-X1) with Co Kα radiation (λ=1.78897Å, 35 kV and 40 mA). Electron backscatter diffraction (EBSD) characterizations were conducted in a Zeiss-Crossbeam focus ion beam scanning electron microscope at 15 kV. Electron channelling contrast imaging (ECCI) characterizations were performed using a Zeiss-Merlin high-resolution field emission electron microscope at 30 kV. Transmission electron microscopy (TEM) analysis including selected area electron diffraction (SAED) was conducted in a JEOL JEM 2100 at 200 kV. Scanning transmission electron microscopy (STEM) images were collected using a probe-corrected Titan Themis 60-300 (Thermo Fischer Scientific) microscope. To modify the Z-contrast characteristics of the imaging mode, high-angle annular dark-field (HAADF) micrographs with a convergence angle of 23.8 mrad were acquired at 300 kV. The resulting collection angle ranges from 73 mrad to 200 mrad. Further energy-dispersive X-ray spectroscopy (EDS) analysis was conducted using Thermo Fischer Scientific's Super-X windowless EDS detector at an acceleration voltage of 300 kV. Atom probe tomography (APT) experiments were performed in a local electrode atom probe (LEAP 5000 XR) from Cameca Instruments Inc and analyzed with commercial AP Suite software (v 6.1). A pulse frequency of 125 kHz, a pulse energy of 40 pJ, and a temperature of 60 K was used. The detection rate was kept at a frequency of 1 ion per 100 pulses.

**Mechanical response measurements.** Room-temperature uniaxial tensile tests were performed using flat tensile specimens at an initial strain rate of $1\times10^{-3}$ s$^{-1}$. The tensile specimens were machined along the rolling direction from the alloy sheets by electrical discharge machining. The specimens with a total length of 60 mm, a gauge length of 30 mm, a gauge width of 5 mm and a thickness of 2 mm were used to probe the bulk tensile properties. Further, smaller tensile specimens with a total length of 20 mm, a gauge length of 10 mm, a gauge width of 2 mm and a thickness of 1 mm were used to measure the local strain evolution by the digital image correlation (DIC) method. At least four specimens for each condition were tested to confirm reproducibility. Further, to clarify the relation between global strain-stress behaviour and microstructure evolution, we also conducted interrupted tensile tests to different global true strains (i.e., 5%, 15% and 25%), and the microstructures in the middle part of the deformed regions were then characterized correspondingly.

**Magnetic response measurements.** The magnetic response was evaluated using the Quantum Design Magnetic Properties Measurement System (MPMS) equipped with a standard vibrating sample magnetometry (VSM) option. Cuboid specimens in 3×3×1 mm$^3$ (*length×width×thickness*) were used for the measurements. The hysteresis loops $M(H)$ were performed in an external magnetic field of ± 800 kA/m at a magnetic field-sweeping rate of 1 kA/m at 10 K, 300 K, 500 K and 800 K, respectively. The temperature dependence of magnetization $M(T)$ analysis was carried out under an applied field of 40 kA/m from 10 K to 1000 K with a temperature-sweeping rate of 10 K/min.

The magnetic domain patterns were characterized by a magneto-optical Kerr effect (MOKE) Zeiss microscope (Axio Imager.D2m). The domain wall movement was captured under an applied magnetic field of ±155 kA/m. Prior to the measurement, a background image was collected as a reference in the AC demagnetized state. The images acquired at different applied fields were enhanced by subtracting the background image using KerrLab software.

**Physical response measurements.** The electrical resistivity response was evaluated using the Quantum Design Physical Properties Measurement System (PPMS) equipped with an Electrical Transport Option (ETO) option. Cuboid specimens in 6×2×1 mm$^3$ (*length×width×thickness*) were used for the measurements. The resistivity $\rho$ values are calculated by:

$$\rho = \frac{RA}{l}$$

where $R$ is the reported resistance, $A$ is the cross-sectional area through which the current is passed and $l$ is the voltage lead separation. The resistance value of each measurement is obtained by averaging those from 100 times of current passing. At least three specimens for each condition were tested.

**Thermodynamic calculations.** The equilibrium composition of the fcc matrix and $L1_2$ particles in the $Co_{28}Fe_{32}Ni_{28}Ta_5Al_7$ (at.%) alloy at 1173 K were calculated using the Thermo-Calc software (v.2022a) equipped with the High Entropy Alloys database TCHEA v.4.2. The calculated equilibrium compositions for the fcc and $L1_2$ phases in the $Co_{28}Fe_{32}Ni_{28}Ta_5Al_7$ (at.%) alloy are $Fe_{36}Co_{31}Ni_{23}Ta_4Al_6$ and $Ni_{63}Ta_{13}Fe_6Co_3Al_{15}$ (at.%), respectively.

**Estimation of particle size (edge-length).** The size distribution is statistically analyzed by applying a batch image processing protocol with multiple 2D projected ECC images of all the MCA samples at different annealed states (Extended Data Fig. 7). The average particle size (edge-length) of the $L1_2$ particles is estimated by:

$$d = \sqrt{\frac{\sum S_i}{i}}$$

where $d$ is the average particle size, $S_i$ is related to the area of each particle acquired from the 2D projected ECC images by batch image processing protocol, $i$ is the total particle number. The particle size of the M-MCA is also characterized by DF-TEM (Fig. 1d) and BF-TEM (Extended Data Fig. 1). The TEM results fit well with the value acquired by ECC images.

**Estimation of interfacial coherency stress.** The coherency stress at the $L1_2$/fcc interface is determined by integrating the lattice misfit across the interface as:

$$\delta = S_{L1_2/fcc} \sum \delta_x$$

where $S_{L1_2/fcc}$ is the $L1_2$/fcc interface area related to the average particle size ($d$), the volume fraction of the $L1_2$ particles ($f$) and the overall volume ($V$) as follows:

$$S_{L1_2/fcc} = \frac{Vf}{d^3} \cdot 6d^2 = \frac{6Vf}{d}$$

$\delta_x$ is the varying lattice misfit as a function of distance ($x$) from the L1$_2$/fcc interface determined by the following equation[18]:

$$\delta_x = 2 \times \left[\frac{a_x^{L1_2} - a_x^{fcc}}{a_x^{L1_2} + a_x^{fcc}}\right]$$

$a_x^{L1_2}$ and $a_x^{fcc}$ are the lattice parameters of the L1$_2$ and fcc phases at the interfacial region. Such values were calculated using the L1$_2$/fcc interfacial chemical compositions acquired from the APT datasets with Vegard's relation[59]:

$$a_x^{L1_2} = a_0^{L1_2} + \sum_i \Gamma_i^{L1_2} x_i^{L1_2}$$

$$a_x^{fcc} = a_0^{fcc} + \sum_i \Gamma_i^{fcc} x_i^{fcc}$$

where $a_0^{L1_2}$ and $a_0^{fcc}$ are the average lattice parameters for the L1$_2$ particles and the fcc matrix, respectively, derived from the Rietveld simulation based on the XRD measurements, as shown in Extended Data Table 1. $\Gamma_i^{L1_2}$ and $\Gamma_i^{fcc}$ are the Vegard coefficients for the L1$_2$ and fcc phases obtained from the ordered Ni$_3$Al phase and disordered fcc phase in the Ni-base superalloys[60], as shown in Extended Data Table 2. Note that the above-calculated lattice misfit $\delta_l$ represents the theoretical unconstrained state. This can be related to the constrained misfit ($\varepsilon$) by elasticity theory as below[61]:

$$\varepsilon = \frac{3}{2} \delta_l$$

The estimated interfacial constrained misfit value is $1.09 \times 10^6$, $4.08 \times 10^5$ and $1.96 \times 10^5$ for the S-MCA, M-MCA and L-MCA, respectively. Therefore, the significant decrease in the interfacial coherency stress is expected to play an essential role in releasing the pining effect on domain wall movement with particle coarsening for the MCAs with the particle size below the domain wall width.

**Estimation of dislocation density.** The dislocation density ($\rho$) in the fcc matrix can be calculated through the Williamson-Smallman relationship as[62]:

$$\rho = \frac{2\sqrt{3}(\varepsilon_s^2)^{1/2}}{Db}$$

where $\varepsilon_s$ is microstrain, $D$ is crystallite size acquired from the XRD profiles (Extended Data Table 1), and $b$ is the Burgers vector (for fcc structure, $b=\sqrt{2}/2 \times a_{\text{fcc}}$)[63]. The dislocation density in the fcc matrix is thus estimated to be $1.50\times10^{14}$ m$^{-2}$, $9.32\times10^{13}$ m$^{-2}$ and $5.38\times10^{13}$ m$^{-2}$ for the S-MCA, M-MCA and L-MCA, respectively. Based on the above estimation, the considerable improvement in the coercivity also derives from the decrease of dislocation density in the fcc matrix.

**Estimation of particle shearing stress.** Based on the experimental observation (Fig. 2c and Extended Data Fig. 2c), particle shearing is the primary deformation mechanism in the investigated MCAs. The strengthening contribution of particle shearing ($\Delta\tau$) is estimated according to[64]:

$$\Delta\tau_{\text{Shearing}} = \frac{F}{b \cdot 2\lambda}$$

where $2\lambda$ is the mean spacing of the particles, $2\lambda \approx \sqrt{\frac{2}{f}} \cdot d$, $d$ is the average particle size, $f$ is the volume fraction of the particles shown in Extended Data Table 1, $F$ is the force exerted on the particles. The shearing strength is expressed as:

$$\Delta\tau_{\text{Shearing}} = k\sqrt{fd}$$

by using the relation $F \propto d^{3/2}$ and introducing constant k. The effect of particle strengthening of the M-MCA is then estimated to be two times larger than that of the S-MCA ($\Delta\tau_{M-MCA}/\Delta\tau_{S-MCA} = \frac{k\sqrt{f_{M-MCA} \cdot r_{M-MCA}}}{k\sqrt{f_{S-MCA} \cdot r_{S-MCA}}}$).

When considering the volume fraction of the particles to be constant, the mean spacing of the particles increases with increasing the particle size. As a result, the force required for shearing particles increases until the Orowan mechanism is activated, i.e., dislocations bowing the particles becomes easier than shearing. The critical mean spacing of the particles is determined by[64]:

$$\Delta\tau_{\text{Shearing}} = \frac{F}{b \cdot 2\lambda} = \Delta\tau_{Orowan} = \frac{Gb}{2\lambda}$$

G=84 GPa is the adopted shear modulus[65]. Consequently, the critical mean spacing of the particles is calculated as 3094.3 nm. However, in the current MCAs, the volume fraction of the L1$_2$ phase is not constant even after annealing at 1173 K for 100 h. This is because the alloys have not yet reached the thermodynamical equilibrium state, as indicated by both thermodynamic calculations and APT analysis (Extended Data Fig. 5).

**Estimation of magnetostatic energy.** The magnetostatic energy ($E_s$) determines the coercive force which interacting between the paramagnetic particles (for M-MCA and L-MCA) and domain wall movement according to the formula[45]:

$$E_s = \frac{1}{2}\mu_0 \frac{1}{3} M_s^2 d^3$$

where $\mu_0 = 4\pi \times 10^{-7}$ H/m is the permeability of vacuum, $d$ is the average particle size, $M_s$ is the saturation magnetization of the fcc matrix. For M-MCA and L-MCA, in which the L1$_2$ phase is paramagnetic (Extended Data Fig. 4), the $M_s$ of the fcc matrix is considered as the overall $M_s$ of the alloy. The values of $E_s$ significantly increase with increasing the particle size, that is, it varies from $1.57 \times 10^{-24}$ (M-MCA) to $3.65 \times 10^{-23}$ (L-MCA). The notable increase in magnetostatic energy results in a strong magnetic pinning effect.

**Estimation of domain wall width.** Strong pinning arises and results in the deterioration of coercivity when the microstructure defects have a comparable dimension to the domain wall thickness ($\delta_w$). As a result, the estimation of the $\delta_w$ helps understand the extremely low coercivity in the current work is given by[66,67]:

$$\delta_w = \pi(A_{ex}/K_1)^{1/2}$$

where $A_{ex} = k_B T_c/2a_0$ is the exchange stiffness, $k_B = 1.380649 \times 10^{-23}$ J/K is the Boltzmann's constant, $T_c$ and $a_0$ are the Curie temperature and lattice parameter of the fcc matrix, respectively (Extended Data Fig. 4d and Extended Data Table 1). $K_1$ is the first magnetocrystalline anisotropy constant. The value of $K_1$ (M-MCA) is taken from the Co-Fe system[68,69] based on the composition of the fcc matrix (Fig. 1f) as 10.4 kJ/m$^3$ (Al and Ta are non-ferromagnetic elements that do not show any magnetic moment, the chemical composition of the fcc phase Fe$_{36}$Co$_{28}$Ni$_{26}$Al$_7$Ta$_3$, at.%, in the M-MCA is thus considered as Co$_{31}$(Fe+Ni)$_{69}$, at.%). The domain wall thickness of the M-MCA is thus estimated to be 112 nm. Similarly, the domain wall thicknesses of the S-MCA and L-MCA are calculated as 103 nm and 117 nm, respectively.

# Acknowledgements

The support of Prof. Gerhard Dehm at the Max-Planck-lnstitut für Eisenforschung is gratefully acknowledged. L. Han would like to acknowledge the financial support from the China Scholarship Council (201906370028). Z. Li. would like to acknowledge the National Natural Science Foundation of China (51971248) and the Hunan Special Funding for the Construction of Innovative Province (2019RS1001). O. Gutfleisch would like to acknowledge the Deutsche Forschungsgemeinschaft (405553726–TRR 270).

**Author Information**

Reprints and permissions information is available at www.nature.com/reprints. The authors declare no competing financial interests. Readers are welcome to comment on the online version of the paper. Correspondence and requests for materials should be addressed to D. Raabe (d.raabe@mpie.de) or Z. Li (lizhiming@csu.edu.cn).